\documentclass[%
reprint,
superscriptaddress,
groupedaddress,
unsortedaddress,
 amsmath,amssymb,
 aps,
]{revtex4-2}

\usepackage{graphicx}
\usepackage{dcolumn}
\usepackage{hyperref}
\usepackage{float}
\usepackage{braket}
\usepackage{color}
\usepackage{subfigure}

\begin{document} 

\title{A group theoretical treatise on color mixing}
\author{Subhankar Roy\footnote{Corresponding author:subhankar@gauhati.ac.in}}
\author{Darshana Chutia\footnote{darsanachutia@gmail.com}}
\author{Abhijit Das \footnote{abhijitdas@gauhati.ac.in}}
\affiliation{%
 Physics Department, Gauhati University
}

             

\begin{abstract}

On relating the three primary colors namely red, green and blue with a triplet representation  of the non-abelian group $SU(3)$, it is established that the three secondary and six tertiary colors along with two neutral colors transform as anti-triplet and octet under the same group respectively. 

\end{abstract}


\maketitle

It is always expected that the fundamental principles of Nature find inherent connections with the group theory. For example, in the simple quark model, the three quarks: up\,($u$), down\,($d$) and strange\,($s$) transform as a $triplet$ ($\mathbf{3}$) or more specifically \textit{flavour-triplet} under the non-abelian group $SU(3)$\,\cite{GellMann:1961ky, GellMann:1962xb, Coleman:1963pj, Hara:1963gw, GellMann:1964nj, Zweig:1981pd, Zweig:1964jf,PETERMANN1965349}. Similarly the three anti-quarks ($\bar{u}$, $\bar{d}$, $\bar{s}$) transform as $SU(3)$  $antitriplet$ ($\mathbf{\bar{3}}$). Based on this fact, the formation and classification of the different families of baryons (the bound states of three quarks) and the mesons (bound state of one quark and one anti-quark) are  explained successfully. Another quantum number \textit{color}\,\cite{PhysRevLett.13.598,PhysRev.139.B1006, Bardeen:1972xk,Fritzsch:1973pi} is introduced in the frame work of quark model to respect the Pauli's exclusion principle. The aforesaid term \textit{color} is no way related to the visual colors and a quark may take anyone of the three color quantum numbers : red {$r$), green ($g$) and blue ($b$). These three color charges form a triplet ($\mathbf{3}$) under $SU(3)$ (\textit{color-triplet}). Similarly the anti quark may take anti-color quantum numbers: $\bar{r}$, $\bar{g}$ and $\bar{b}$ which transform as \textit{color anti-triplet} ($\bar{3}$) under $SU(3)$. As the baryons and the mesons do not show direct evidence of color quantum number, the possible bound states of quarks or\,(and) antiquarks are assumed as color-neutral or  \textit{color-singlet} under $SU(3)$. The quark model is a quantum mechanical framework and inspired by the success of simple quark model, the present work probes to investigate the significance and applicability of $SU(3)$ group in the context of \textit{visual} colors. We see that in case of the quark model, the three quarks serve as the fundamental building blocks. Similarly, in color mixing phenomenon, there exist three primary colors\,\cite{johnston2015history} namely, Red (R), Green (G) and Blue (B). When two different primary colors are mixed at a time, one gets three secondary colors namely, \textit{yellow}, \textit{magenta} and \textit{cyan} as outcome\, \cite{rainwaterlight}. This possibility is excluded in quark model as we never come across a bound state of two quarks. The secondary entries when mixed with the individual primary colors, result in a set of six tertiary colors comprising of \textit{chartreuse}, \textit{orange}, \textit{rose}, \textit{violet}, \textit{azure}, and \textit{spring green}\cite{bendito2005rgb}. The tertiary color states are akin to the baryonic states in the quark model. 

In this model, we consider a three-dimensional \textit{visual color vector space} ($V^{3}$) with three primary colors: R, G and B serving as the basis vectors, and also we assume that the basis,
\begin{eqnarray}
\begin{pmatrix}
R\\
G\\
B
\end{pmatrix},\nonumber
\end{eqnarray}
transform as a triplet $\mathbf{3}$ under the $SU(3)$ group. We mark this triplet ($\mathbf{3}$)  as \textit{Visual triplet}. To simplify the discussion, we first take over the context of $SU(2)$ group which happens to be a subgroup of $SU(3)$. For the former, the fundamental irreducible representation is a \textit{doublet}. So, in this framework, we take a simplified two dimensional visual color vector space ($V^2$) which is spanned by $R$ and $G$ only. Let us consider that the vector,
\begin{eqnarray}
\begin{pmatrix}
R\\
G
\end{pmatrix},\nonumber
\end{eqnarray}
transforms as the \textit{visual doublet} $\mathbf{2}$ under $SU(2)$. 

The formation of a secondary color state by mixing the two primaries $R$ and $G$ is to be understood as a \textit{direct product} of the two independent color vector spaces: $V^2_1\otimes V^2_2$ which results in two possibilities: antisymmetric singlet ($\mathbf{1_{A}}$) or a color state belonging to a family of symmetric triplet ($\mathbf{3_{S}}$),
\begin{eqnarray}
2\otimes 2= 1_{A}\oplus 3_{S}.
\end{eqnarray}
The visual singlet $\mathbf{1_{A}}$ corresponds to a single visual color state yellow:
\begin{eqnarray}
\ket{yellow} &=& \frac{1}{\sqrt{2}}(R_1 G_2-G_1 R_2),\nonumber
\end{eqnarray}
whereas, the $\mathbf{3_{S}}$ carries three visual color states with yellow as one of the members,
\begin{eqnarray}
\ket{x} &=& R_1 R_2 \nonumber\\
\ket{yellow} &=& \frac{1}{\sqrt{2}}(R_1 G_2+G_1 R_2),\nonumber\\
\ket{y} &=& G_1 G_2.\nonumber
\end{eqnarray}
So, one sees that the color state \textit{yellow} can either be $\mathbf{1_A}$ or may belong to the $\mathbf{3_{S}}$ family. It is realized that if two same primary colors\,(say, Red) were combined, it rather results in the original primary color\,(red) than a secondary. Thus, the members $x$ and $y$ of $\mathbf{3_{S}}$ family are none other than the primary color states R and G respectively. Hence, $\mathbf{3_{S}}$ does not represent a physically meaningful family of secondary colors and the \textit{yellow} is to be marked as a visual color singlet state under $SU(2)$. This discussion embarks the following \textbf{exclusion rules},
\begin{itemize}
\item \textit{The direct product states with all entries same are unphysical.} 

The states namely, $RR$, $RRR$, $RRRR...$ are forbidden. 
\item\textit{ Certain family carrying at least an unphysical state is forbidden.}
\end{itemize}

This exclusion rule is necessary in order to avoid certain unphysical outcome that may arise in further analysis.   

The generation of the tertiary color states corresponds to $V^2_1\otimes V^2_2\otimes V^2_3$. Under $SU(2)$, such combination results in two doublets and one quartet families,
\begin{eqnarray}
2\otimes 2\otimes 2 &=& 2_{M_{A}}\oplus 2_{M_{S}}\oplus 4_{S}.
\end{eqnarray} 

As per the exclusion rules, the family $4_S$ is knocked out. We obtain orange and chartreuse as secondary color states which together transform either as $\mathbf{2_{M_{A}}}$ or $\mathbf{2_{M_{S}}}$. 

The $SU(2)$ group motivates for a triplet representation also. Hence, it opens a possibility to consider,
\begin{eqnarray}
\begin{pmatrix}
R\\
G\\
B
\end{pmatrix},\nonumber
\end{eqnarray}
to transform as visual triplet under $SU(2)$ group. The formulation of the secondary and tertiary colors states corresponds to the following multiplication rules,
\begin{eqnarray}
\label{yy}
3\otimes 3 &=& 5 \oplus 3\oplus 1,
\end{eqnarray}
and
\begin{eqnarray}
\label{tt}
3\otimes 3\otimes 3 &=& 7\oplus 5\oplus 5\oplus 3\oplus 3\oplus 3\oplus 1,
\end{eqnarray}
respectively. It is found that the three secondary colors: \textit{yellow}, \textit{magenta} and \textit{cyan} form a triplet as shown in the right side of eq.\,(\ref{yy}). But due to the exclusion rules, the six tertiary colors are unfounded in any of the multiplet families on the right side of eq.\,(\ref{tt}). Hence it seems that the triplet representation of the three primary colors under $SU(2)$ does not lead to a physically meaningful outcome. 

A promising framework is expected in the light of $SU(3)$ group. As per the group algebra of $SU(3)$ is concerned, the direct product of the two independent primary color vector spaces ($V^3_1 \otimes V^3_2$), or product of two visual color triplets results in a sextet family ($\mathbf{6_S}$) and an anti-visual color triplet ($\mathbf{\bar{3}_A}$),
\begin{eqnarray}
3\otimes 3 = 6_s \oplus \bar{3}_A.
\end{eqnarray}
The first family is forbidden as per the selection rules are concerned. But interestingly the $\mathbf{\bar{3}_A}$ accommodates all three secondary color states
\begin{eqnarray}
\ket{yellow}&=&\frac{1}{\sqrt{2}} (R_1 G_2 -G_1 R_2),\nonumber\\
\ket{magenta}&=&\frac{1}{\sqrt{2}} (R_1 B_2 -B_1 R_2),\nonumber\\
\ket{cyan}&=&\frac{1}{\sqrt{2}} (G_1 B_2 -B_1 G_2),\nonumber
\end{eqnarray} 
and these color states serve as bases of this three-dimensional sub-vector space. Here, we wish to emphasize that in contrast to the present scenario, the $3\otimes 3 $ is forbidden in the quark model as we never encounter a bound state of two quarks. Further, we see that the product of three independent color triplets results in one singlet, two octets and one decuplet family.
\begin{eqnarray}
3\otimes 3\otimes 3 &=& 1_{A}\oplus 8_{M_{S}} \oplus 8_{M_{A}}\oplus 10_{S}.
\end{eqnarray}  
We see that the exclusion principle rejects the decuplet\,($\mathbf{10_{S}}$) family and the six tertiary colors can be fitted well with any one of the octet families: $\mathbf{8_{M_{S}}}$ or $\mathbf{8_{M_{A}}}$. To exemplify, let us consider the states of the $\mathbf{8_{M_{S}}}$ family.
\begin{eqnarray}
\ket{orange} &=& \frac{1}{\sqrt{6}}\left(2\,R_1 R_2 G_3 -R_1 G_2 R_3-G_1 R_2 R_3 \right),\nonumber\\
\ket{chartreuse} &=& \frac{-1}{\sqrt{6}}\left(2G_1 G_2 R_3 + G_1 R_2 G_3 + G_1 R_2 G_3 \right),\nonumber\\
\ket{rose} &=&\frac{1}{\sqrt{6}}\left(2 R_1 R_2 B_3 -R_1 B_2 R_3-B_1 R_2 R_3 \right),\nonumber\\
\ket{X} &=& \frac{1}{\sqrt{12}}\left( 2 R_1 G_2 B_3-R_1 B_2 G_3-G_1 B_2 R_3\right.\nonumber\\
&& \left. +2 G_1 R_2 B_3-B_1 R_2 G_3-B_1 G_2 R_3\right)\nonumber\\
\ket{spring green} &=& \frac{1}{\sqrt{6}}\left(2 G_1 G_2 B_3 -G_1 B_2 G_3-B_1 G_2 G_3 \right),\nonumber\\
\ket{Y} &=& \frac{1}{2}\left(R_1 B_2 G_3+B_1 R_2G_3-B_1 G_2 R_3\right. \nonumber\\
&& \left.-G_1 B_2 R_3\right), \nonumber\\
\ket{violet} &=& \frac{1}{\sqrt{6}} \left(B_1 R_2 B_3 + R_1 B_2 B_3 -2 B_1 B_2 R_3\right), \nonumber\\
\ket{azure} &=& \frac{1}{\sqrt{6}} \left(B_1 G_2 B_3 + G_1 B_2 B_3 -2 B_1 B_2 G_3\right).\nonumber
\end{eqnarray}

We see that there are two color states \textbf{X} and \textbf{Y} having the same color content\,($RGB$) in the above description. At first look, it appears that both \textbf{X} and \textbf{Y} are identical and they resemble the same color state. But as we know, that $SU(2)$ is a subgroup of $SU(3)$, the members of $\mathbf{8_{M_{s}}}$ are treated in a different way under $SU(2)$. The  tertiary sates:\textit{orange} and \textit{chartreuse} form a $SU(2)$ visual doublet and same is true for the \textit{violet} and \textit{azure} pair. Under $SU(2)$ \textbf{X} falls in a triplet family along with other members \textit{rose}, and \textit{spring green}. whereas the color state \textbf{Y} behaves as $SU(2)$ singlet. This shows that the \textbf{X} and \textbf{Y} though appear similar, are different in deed. Let us try to understand the difference between these two states physically following a thought experiment. We consider a set-up where white light from three different sources after passing through perfect  R, G and B filters are made to overlap over a certain surface, say, \textit{\textbf{S}}. So, what resultant color the eye senses will depend not only on the color content but also on the nature of \textit{\textbf{S}}. If \textit{\textbf{S}} is perfectly reflecting, then the eyes sense it as white and for a perfectly absorbing \textit{\textbf{S}}, we see black. Hence, the same combination of R, G and B may appear either as white or black depending on the nature of the surface. So, we identify the $SU(2)$ triplet state $\ket{X}$ as white and the singlet state $\ket{Y}$ as black. But for the sake of argument, the surface \textit{\textbf{S}} could have been designed in such  a way that it can reflect B and G but absorbs R. So in this case we shall not see black. But the eyes will sense only the presence of B and G (\textit{yellow}). Therefore, this mixture is rather secondary than tertiary. Hence this goes against the context of $3\otimes 3\otimes 3$. Similarly, if \textit{\textbf{S}} absorbs two colors, there will be no mixing, and the state will be primary. We can conclude that the six tertiary colors along with two neutral colors white and black complete an octet family under $SU(3)$. In contrast, the \textit{color} quantum number\,(from particle physics), similar to the visual colors also transforms as $\mathbf{3}$ under $SU(3)$ and baryons which is a three quark state takes only the $\mathbf{1_{A}}$ as color wave-function than the octets or decuplets. 

We conclude that continuous group like $SU(2)$ and $SU(3)$ play a promising role in understanding the theory of color mixing. All living beings do not visualize the world in the same way. For example, the human sense is attributed only to three color filters namely red, green and blue and this motivated us to adopt the triplet representation of $SU(3)$. Similarly, for other living creatures, the choice of group is expected to be different. To simplify the discussion, only up to the level of tertiary color generation is concentrated. But we believe that further generation and categorization of the colors can be studied in the same line.

\section*{Acknowledgments}

One of the authors SR thanks N.N Singh at Manipur University, India for the useful discussion. 

\section*{Author contributions}

All authors contributed extensively to this work. DC came up with the problem and did the mathematical exercise. AD contributed to the understanding of color mixing. SR designed the model, interpreted the mathematical results and prepared the manuscript. 

\section*{Competing interests}

The authors declare no competing interests.


%

\end{document}